\documentclass[longauth]{aa}
\usepackage{graphicx}
\usepackage{txfonts}
\usepackage{xcolor}
\usepackage{hyperref}
\hypersetup{colorlinks, linkcolor=blue, citecolor=blue, urlcolor=blue}

\newcommand{\te}{t_{\rm E}}
\newcommand{\thetae}{\theta_{\rm E}}

\newcommand{\pie}{\pi_{\rm E}}
\newcommand{\pirel}{\pi_{\rm rel}}

\newcommand{\dl}{D_{\rm L}}
\newcommand{\ds}{D_{\rm S}}

\definecolor{brown}{rgb}{0.59, 0.29, 0.0}
\definecolor{darkgreen}{rgb}{0.0, 0.42, 0.24}
\definecolor{darkblue}{rgb}{0.01, 0.31, 0.59}
\definecolor{darkblue}{rgb}{0.0, 0.25, 0.42}
\definecolor{blue}{rgb}{0.0,0.0,1.0}
\definecolor{green}{rgb}{0.0,1.0,0.0}

\def\eqalign#1{\null\,\vcenter{\openup\jot
        \ialign{\strut\hfil$\displaystyle{##}$&$
        \displaystyle{{}##}$\hfil \crcr#1\crcr}}\,}

\begin{document}

\title{KMT-2021-BLG-1547Lb: Giant microlensing planet detected through a signal deformed by source binarity }
\titlerunning{KMT-2021-BLG-1547Lb}

\author{
     Cheongho~Han\inst{01}
\and Weicheng~Zang\inst{02,03}
\and Youn~Kil~Jung\inst{04,05}
\and Ian~A.~Bond\inst{06}
\\
(Leading authors)\\
     Sun-Ju~Chung\inst{02,04}
\and Michael~D.~Albrow\inst{07}
\and Andrew~Gould\inst{08,09}
\and Kyu-Ha~Hwang\inst{04}
\and Yoon-Hyun~Ryu\inst{04}
\and In-Gu~Shin\inst{02}
\and Yossi~Shvartzvald\inst{10}
\and Hongjing~Yang\inst{03}
\and Jennifer~C.~Yee\inst{02}
\and Sang-Mok~Cha\inst{04,11}
\and Doeon~Kim\inst{01}
\and Dong-Jin~Kim\inst{04}
\and Seung-Lee~Kim\inst{04}
\and Chung-Uk~Lee\inst{04}
\and Dong-Joo~Lee\inst{04}
\and Yongseok~Lee\inst{04,11}
\and Byeong-Gon~Park\inst{04}
\and Richard~W.~Pogge\inst{09}
\\
(The KMTNet collaboration)\\
     L. A. G. Monard\inst{12}
\and Qiyue Qian\inst{03}
\and Zhuokai Liu\inst{13}
\and Dan Maoz\inst{14}
\and Matthew T. Penny\inst{15} 
\and Wei Zhu\inst{03}
\\
(The MAP and $\mu$FUN Follow-up Team)\\
     Fumio~Abe\inst{16}
\and Richard~Barry\inst{17}
\and David~P.~Bennett\inst{17,18}
\and Aparna~Bhattacharya\inst{17,18}
\and Hirosame~Fujii\inst{16}
\and Akihiko~Fukui\inst{19,20}
\and Ryusei~Hamada\inst{21}
\and Yuki~Hirao\inst{22}
\and Stela~Ishitani Silva\inst{17,23}
\and Yoshitaka~Itow\inst{16}
\and Rintaro~Kirikawa\inst{21}
\and Iona~Kondo\inst{21}
\and Naoki~Koshimoto\inst{24}
\and Yutaka~Matsubara\inst{16}
\and Shota~Miyazaki\inst{25}
\and Yasushi~Muraki\inst{16}
\and Greg~Olmschenk\inst{17}
\and Cl{\'e}ment~Ranc\inst{26}
\and Nicholas~J.~Rattenbury\inst{27}
\and Yuki~Satoh\inst{21}
\and Takahiro~Sumi\inst{21}
\and Daisuke~Suzuki\inst{21}
\and Mio~Tomoyoshi\inst{21}
\and Paul~J.~Tristram\inst{28}
\and Aikaterini~Vandorou\inst{17,18}
\and Hibiki~Yama\inst{21}
\and Kansuke~Yamashita\inst{21}
\\
(The MOA Collaboration)\\
}

\institute{
      Department of Physics, Chungbuk National University, Cheongju 28644, Republic of Korea,                                                            
\and  Center for Astrophysics $|$ Harvard \& Smithsonian, 60 Garden St., Cambridge, MA 02138, USA                                                        
\and  Department of Astronomy, Tsinghua University, Beijing 100084, China                                                                                
\and  Korea Astronomy and Space Science Institute, Daejon 34055, Republic of Korea                                                                       
\and  Korea University of Science and Technology, Korea, (UST), 217 Gajeong-ro, Yuseong-gu, Daejeon, 34113, Republic of Korea                            
\and  Institute of Natural and Mathematical Science, Massey University, Auckland 0745, New Zealand                                                       
\and  University of Canterbury, Department of Physics and Astronomy, Private Bag 4800, Christchurch 8020, New Zealand                                    
\and  Max-Planck-Institute for Astronomy, K\"{o}nigstuhl 17, 69117 Heidelberg, Germany                                                                   
\and  Department of Astronomy, Ohio State University, 140 W. 18th Ave., Columbus, OH 43210, USA                                                          
\and  Department of Particle Physics and Astrophysics, Weizmann Institute of Science, Rehovot 76100, Israel                                              
\and  School of Space Research, Kyung Hee University, Yongin, Kyeonggi 17104, Republic of Korea                                                          
\and  Klein Karoo Observatory, Calitzdorp, and Bronberg Observatory, Pretoria, South Africa                                                              
\and  Kavli Institute for Astronomy and Astrophysics, Peking University, Yi He Yuan Road 5, Hai Dian District, Beijing 100871, China                     
\and  School of Physics and Astronomy, Tel-Aviv University, Tel-Aviv 6997801, Israel                                                                     
\and  Department of Physics and Astronomy, Louisiana State University, Baton Rouge, LA 70803, USA                                                        
\and  Institute for Space-Earth Environmental Research, Nagoya University, Nagoya 464-8601, Japan                                                        
\and  Code 667, NASA Goddard Space Flight Center, Greenbelt, MD 20771, USA                                                                               
\and  Department of Astronomy, University of Maryland, College Park, MD 20742, USA                                                                       
\and  Department of Earth and Planetary Science, Graduate School of Science, The University of Tokyo, 7-3-1 Hongo, Bunkyo-ku, Tokyo 113-0033, Japan      
\and  Instituto de Astrof{\'i}sica de Canarias, V{\'i}a L{\'a}ctea s/n, E-38205 La Laguna, Tenerife, Spain                                               
\and  Department of Earth and Space Science, Graduate School of Science, Osaka University, Toyonaka, Osaka 560-0043, Japan                               
\and  Institute of Astronomy, Graduate School of Science, The University of Tokyo, 2-21-1 Osawa, Mitaka, Tokyo 181-0015, Japan                           
\and  Department of Physics, The Catholic University of America, Washington, DC 20064, USA                                                               
\and  Department of Astronomy, Graduate School of Science, The University of Tokyo, 7-3-1 Hongo, Bunkyo-ku, Tokyo 113-0033, Japan                        
\and  Institute of Space and Astronautical Science, Japan Aerospace Exploration Agency, 3-1-1 Yoshinodai, Chuo, Sagamihara, Kanagawa 252-5210, Japan     
\and  Sorbonne Universit\'e, CNRS, UMR 7095, Institut d'Astrophysique de Paris, 98 bis bd Arago, 75014 Paris, France                                     
\and  Department of Physics, University of Auckland, Private Bag 92019, Auckland, New Zealand                                                            
\and  University of Canterbury Mt.~John Observatory, P.O. Box 56, Lake Tekapo 8770, New Zealand                                                          
}


\abstract
{}
{
We investigate the previous microlensing data collected by the KMTNet survey in search of
anomalous events for which no precise interpretations of the anomalies have been suggested.
From this investigation, we find that the anomaly in the lensing light curve of the event
KMT-2021-BLG-1547 is approximately described by a binary-lens (2L1S) model with a lens
possessing a giant planet, but the model leaves unexplained residuals.
}
{
We investigate the origin of the residuals by testing more sophisticated models that include
either an extra lens component (3L1S model) or an extra source star (2L2S model) to the 2L1S
configuration of the lens system. From these analyses, we find that the residuals from the 2L1S
model originate from the existence of a faint companion to the source. The 2L2S solution
substantially reduces the residuals and improves the model fit by $\Delta\chi^2=67.1$ with 
respect to the 2L1S solution. The 3L1S solution also improves the fit, but its fit is worse 
than that of the 2L2S solution by $\Delta\chi^2=24.7$.
}
{
According to the 2L2S solution, the lens of the event is a planetary system with planet and
host masses $(M_{\rm p}/M_{\rm J}, M_{\rm h}/M_\odot)=\left( 1.47^{+0.64}_{-0.77}, 
0.72^{+0.32}_{-0.38}\right)$ lying at a distance $\dl =5.07^{+0.98}_{-1.50}$~kpc, and the 
source is a binary composed of a subgiant primary of a late G or an early K spectral type 
and a main-sequence companion of a K spectral type.  The event demonstrates the need of 
sophisticated modeling for unexplained anomalies for the construction of a complete 
microlensing planet sample. 
}
{}


\maketitle

\section{Introduction}\label{sec:one}

The planetary signal in a lensing light curve is mostly described by a 2L1S model, in which 
the lens comprises two masses of the planet and its host and the source is a single star 
\citep{Mao1991, Gould1992}. It occasionally happens that a planetary signal cannot be
precisely described by the usual 2L1S model because of several major causes.

The first cause of the deviation of a planetary signal from a 2L1S form is the existence of 
an additional planet. In general, a planet induces two sets of caustics, in which one lies 
near the position of the planet host (central caustic), and the other lies away from the host 
(planetary caustic) at the position ${\bf s}-1/{\bf s}$, where ${\bf s}$ denotes the position 
vector of the planet from the host \citep{Griest1998, Han2006}.  For a lens system containing 
multiple planets, the central caustics induced by the individual planets appear in a common 
region around the planet host, and thus the magnification pattern of the central region deviates 
from that of a single-planet system \citep{Gaudi1998}, causing deformation of the planetary 
signal.  There have been five cases of microlensing events with planetary signals deformed by 
multiple planets, including OGLE-2006-BLG-109 \citep{Gaudi2008, Bennett2010}, OGLE-2012-BLG-0026 
\citep{Han2013, Beaulieu2016}, OGLE-2018-BLG-1011 \citep{Han2019}, OGLE-2019-BLG-0468 
\citep{Han2022d}, and KMT-2021-BLG-1077 \citep{Han2022a}.

The second cause for the deformation of a planetary signal is the binarity of the planet 
host.  Under the lens configuration in which a planet orbits around one component of a wide 
binary star or around the barycenter of a close binary star, the binary companion induces 
additional perturbations in the central magnification region, and thus the signal of the 
planet may deviate from a single-planet form \citep{Lee2008}. There have five reports of 
microlensing planets with signals affected by binary companions to the hosts, including 
OGLE-2006-BLG-284 \citep{Bennett2020}, OGLE-2007-BLG-349 \citep{Bennett2016}, OGLE-2016-BLG-0613 
\citep{Han2017}, OGLE-2018-BLG-1700 \citep{Han2020}, and KMT-2020-BLG-0414 \citep{Zang2021}.

A planetary signal can also be deformed by the binarity of the source star. If the source 
is accompanied by a close companion, the perturbation region induced by the planet can be
additionally swept by the companion star to the primary source, and this can induce a
deformation of the planetary signal. There have been three cases of planetary signals that 
were affected by the existence of the source companions, including MOA-2010-BLG-117 
\citep{Bennett2018}, KMT-2018-BLG-1743 \citep{Han2021a}, and KMT-2021-BLG-1898 
\citep{Han2022b}.

It is known that firmly identifying the cause of the deformation in a planetary signal is 
often difficult as demonstrated in the case of KMT-2021-BLG-0240. For this event, 
\citet{Han2022c} found that the central anomaly could be explained with either a triple-lens 
(3L1S) model, in which the lens is composed of three masses including two planets and their 
host, or a binary-lens binary-source (2L2S) model, in which the lens is a single planet system 
and the source is a binary.

We have conducted systematic investigation of the microlensing data collected in previous 
seasons by the Korea Microlensing Telescope Network \citep[KMTNet:][]{Kim2016} survey in 
search of anomalous lensing events, for which no precise interpretations of the anomalies 
have been suggested. From this investigation, \citet{Han2023a} found that the precise 
descriptions of the anomalies in the two lensing events OGLE-2018-BLG-0584 and 
KMT-2018-BLG-2119 required four-body (lens plus source) lensing models, in which both the 
lens and source are binaries.  \citet{Han2023b} also found that the description of the 
anomaly that appeared in the lensing event KMT-2021-BLG-1122 required a different four-body 
lensing model, in which the lens is a triple stellar system and the source is a single star. 
In this work, we present the analysis of the lensing event KMT-2021-BLG-1547, which was left 
without a suggested lensing solution that precisely describes the anomaly appearing in the 
lensing light curve.

For the presentation of the analysis, we organize the paper as follows.  In Sect.~\ref{sec:two}, 
we describe the observations of the lensing event, instrument used for observations, and the 
reduction procedure of the data. In Sect.~\ref{sec:three}, we depict the analysis of the 
observed lensing light curve under various models of the lens-system configurations, including 
2L1S (in Sect.~\ref{sec:three-one}), 3L1S (Sect.~\ref{sec:three-two}), and 2L2S 
(Sect.~\ref{sec:three-three}) models.  In Sect.~~\ref{sec:four}, we specify the source of the 
event and estimate the angular Einstein radius of the lens system. In Sect.~\ref{sec:five}, 
we present the physical parameters of the lens system estimated from the Bayesian analysis 
of the lensing event.  In Sect.~\ref{sec:six}, we summarize the results found from the analysis 
and conclude.

\begin{figure}[t]
\includegraphics[width=\columnwidth]{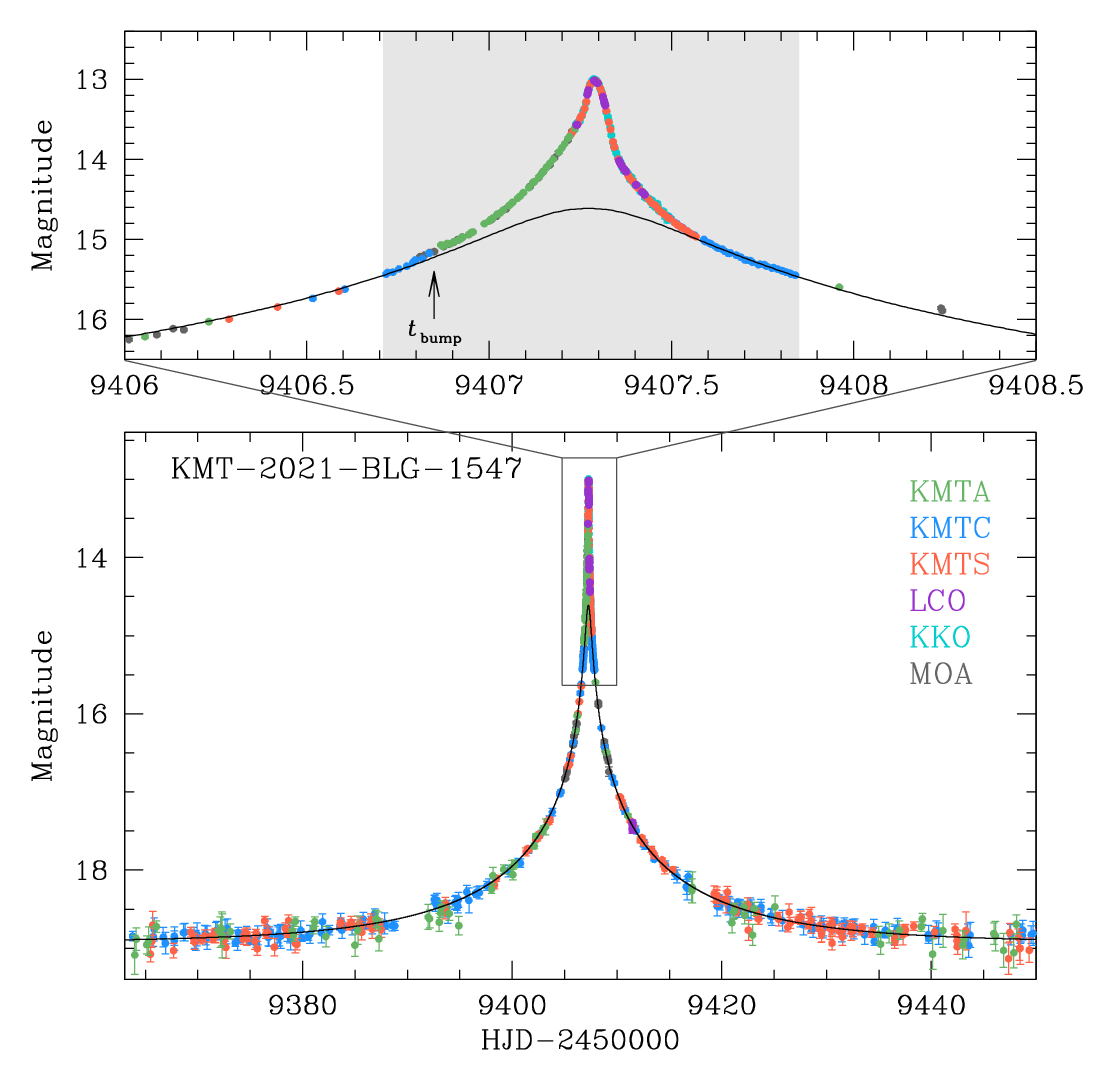}
\caption{
Light curve of the microlensing event KMT-2021-BLG-1547. The lower and upper panels show the 
whole view and enlarged view around the anomalous region near the peak, respectively. The 
solid curve drawn over the data point is a 1L1S model obtained by excluding the data around 
the the anomaly region of the light curve. The shaded region in the upper panel represents 
the duration of intensive observations, and the arrow indicates the position of a weak bump. 
}
\label{fig:one}
\end{figure}

\begin{figure*}[t]
\centering
\includegraphics[width=12.0cm]{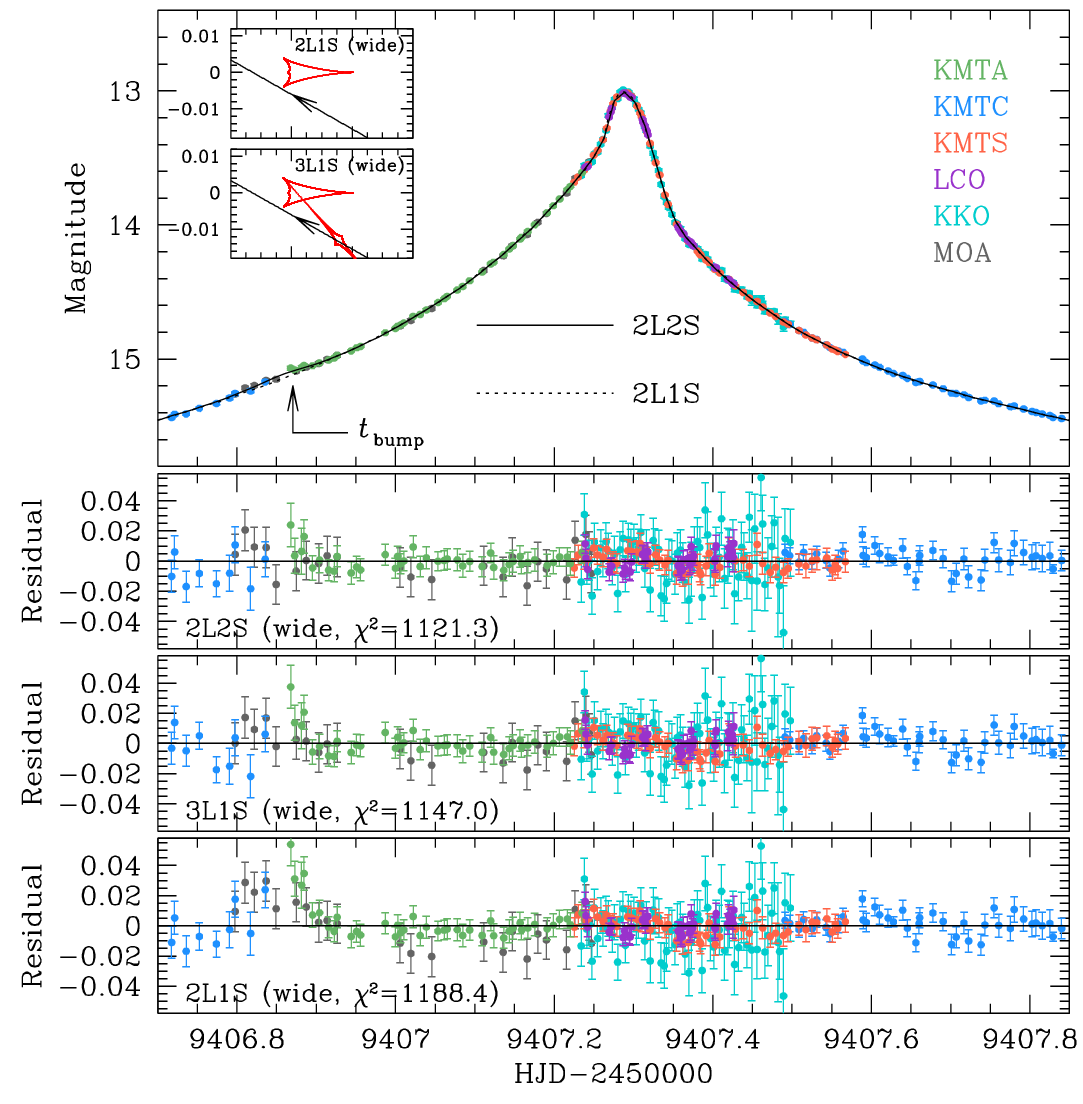}
\caption{
Zoom-in view around the peak region of the KMT-2021-BLG-1547 light curve. The lower three 
panels show the residuals from the 2L2S (wide), 3L1S (wide), and 2L1S (wide) models. The 
dotted and solid curves drawn over the data points in the top panel are the models of the 
2L1S and 2L2S solutions, respectively. The arrow marked by "$t_{\rm bump}$" indicates the 
region that leaves a bump in the residual from the 2L1S model. The two insets in the top 
panel show the lens-system configurations of the 2L1S and 3L1S models. In each inset, the 
red figures are the caustics, and the line with an arrow represents the source trajectory. 
}
\label{fig:two}
\end{figure*}

\section{Observation and data}\label{sec:two}

The source of the lensing event KMT-2021-BLG-1547 lies toward the Galactic bulge field at 
the equatorial coordinates $({\rm RA}, {\rm DEC})_{\rm J2000} = $ (18:09:35.90, -29:05:02.18), 
which correspond to the Galactic coordinates $(l, b) = (2^\circ\hskip-2pt .494, 
-4^\circ\hskip-2pt .614)$. The baseline magnitude of the source is $I_{\rm base} = 19.09$, 
and the extinction toward the field is $A_I = 0.83$.  The lensing-induced magnification of 
the source flux was first found by the KMTNet group on 2021 July 1, which corresponds to the 
abridged Heliocentric Julian Date ${\rm HJD}^\prime \equiv {\rm HJD}-2450000 = 9396.5$, when 
the source became brighter than the baseline magnitude by $\Delta I\sim 0.63$~mag. The position 
of the source corresponds to the KMTNet BLG33 field, toward which observations in a normal 
survey mode were conducted with a 2.5~hr cadence. On 2021 July 7, ${\rm HJD}^\prime \sim 9402$, 
the event was independently found by the Microlensing Observations in Astrophysics survey 
(MOA) group \citep{Bond2001}, who referred to the event as MOA-2021-BLG-228. We hereafter 
designate the event as KMT-2021-BLG-1547 in accordance with the convention of the microlensing 
community using the event ID reference of the first discovery survey. The KMTNet observations 
of the event were done with the use of three identical 1.6~m telescopes, that are distributed 
at three sites of the Southern Hemisphere: the Siding Spring Observatory (SSO) in Australia 
(KMTA), the Cerro Tololo Interamerican Observatory (CTIO) in Chile (KMTC), and the South African 
Astronomical Observatory (SAAO) in South Africa (KMTS).  The MOA observations were done utilizing 
the 1.8~m telescope of the Mt. John Observatory in New Zealand.

Images containing the source of the event were mostly acquired in the $I$ band for the KMTNet
survey and in the customized MOA-$R$ band for the MOA survey. The initial reduction of the
images and photometry of the source were done using the pipelines developed by \citet{Albrow2009} 
and \citet{Bond2001} for the KMTNet and MOA surveys, respectively.  For the optimization of the 
data, the KMTNet data used in the analysis were prepared by rereducing the images using the 
updated TLC algorithms developed by \citet{Yang2023}.  In order to set the scatter of the data 
to be consistent with the error bars and to set $\chi^2$ per degree of freedom (dof) for each 
data set to unity, we readjusted the error bars of the data resulting from the automatized 
photometry pipelines using the \citet{Yee2012} routine. For a subset of KMTC images taken in 
$V$ and $I$ bands, we conducted an additional photometry using the pyDIA code of \citet{Albrow2017} 
for the source color measurement.

The lensing light curve of KMT-2021-BLG-1547 constructed with the combined KMTNet and MOA data 
is presented in Figure~\ref{fig:one}, in which the lower panel shows the whole view and the 
upper panel shows the zoom-in view of the peak region. The light curve peaked at ${\rm HJD}^\prime 
= 9407.280$ with a very high magnification of $A_{\rm max}\sim 299$.  Because the peak region 
of a very high-magnification event is sensitive to planetary signals \citep{Griest1998}, an 
alert was issued by the KMTNet HighMagFinder system \citep{Yang2022} to cover the peak region 
of the light curve.  In response to this alert, the KMTNet group increased its observational 
cadence to $\sim 0.15$~hr, which is about 17 times shorter than the cadence of the normal 
survey mode. Furthermore, the MAP \& $\mu$FUN Follow-up teams \citep{Zang2021} carried out 
followup observations of the event around the peak of the lensing light curve using the 1.0~m 
telescope of Las Cumbres Observatory (LCO) at SAAO and the 0.36~m telescope of Klein Karoo 
Observatory (KKO) in South Africa. The KKO data were acquired at a very high cadence, and we 
use a binned data with a 5~min interval. The densely covered peak region, the shaded region 
in the upper panel of Figure~\ref{fig:one}, revealed a clear signature of an anomaly, that 
lasted slightly less than a day. We note that the high-magnification alert was issued at UT 
20:15 on 2021 July 10 (${\rm HJD}^\prime =9406.34$), which was well before the anomaly was 
noticed.  In Figure~\ref{fig:two}, the magnitudes of the data collected from MOA and MAP \& 
$\mu$FUN Follow-up observations are scaled to the KMTNet system by linearly aligning the source 
flux to that  of the KMTNet data.

\section{Light curve analysis}\label{sec:three}

\subsection{2L1S model}\label{sec:three-one}

From pattern of the anomaly, in which the light curve rapidly rises and falls during a short 
period of time, it is likely that a caustic is involved in the pattern of the anomaly. 
Therefore, we began analysis by modeling the light curve under a 2L1S interpretation. The 
modeling was carried out in search of a lensing solution, which indicates a set of the lensing 
parameters describing the light curve. Under the approximation of the rectilinear relative 
motion between the lens and source, a 2L1S lensing light curve is depicted by 7 parameters, 
including $t_0$, $u_0$, $\te$, $s$, $q$, $\alpha$, and $\rho$.  The first three parameters 
$(t_0, u_0, \te)$ describe the lens--source approach, and the individual parameters denote 
the time of the closest lens-source approach, the lens-source separation (normalized to the 
angular Einstein radius $\thetae$) at $t_0$, and the event time scale defined as the time 
required for the source to cross $\thetae$. The next three parameters $(s, q, \alpha)$ are 
related to the binary lens, and they indicate the projected separation (scaled to $\thetae$) 
and mass ratio between the binary-lens components ($M_1$ and $M_2$), and the angle of the 
source trajectory with respect to the $M_1$--$M_2$ axis, respectively.  The last parameter 
$\rho$, defined as the ratio of the angular source radius $\theta_*$ to $\thetae$, depicts 
the parts of the lensing light curve affected by finite-source effects.

\begin{table}[t]
\small
\caption{Model parameters of 2L1S solutions\label{table:one}}
\begin{tabular*}{\columnwidth}{@{\extracolsep{\fill}}lcccc}
\hline\hline
\multicolumn{1}{c}{Parameter}    &
\multicolumn{1}{c}{Close}        &
\multicolumn{1}{c}{Wide}        \\
\hline
$\chi^2$/dof            &   $1248.5             $  &  $1188.4             $    \\
$t_0$ (HJD$^\prime$)    &   $9407.280 \pm 0.001 $  &  $9407.278 \pm 0.001 $    \\
$u_0$ ($10^{-3}$)       &   $5.57 \pm 0.10      $  &  $5.45 \pm 0.12      $    \\
$\te$ (days)            &   $19.62 \pm 0.33     $  &  $20.35 \pm 0.40     $    \\
$s  $                   &   $0.726 \pm 0.001    $  &  $1.372 \pm 0.002    $    \\
$q  $ ($10^{-3}$)       &   $2.15 \pm 0.04      $  &  $2.11 \pm 0.05      $    \\
$\alpha$ (rad)          &   $0.455 \pm 0.002    $  &  $0.443 \pm 0.003    $    \\
$\rho$ ($10^{-3}$)      &   $1.55 \pm 0.03      $  &  $1.49 \pm 0.03      $    \\
\hline
\end{tabular*}
\tablefoot{ ${\rm HJD}^\prime = {\rm HJD}- 2450000$.  }
\end{table}

The 2L1S modeling was conducted in two stages. In the first stage, we divided the lensing parameters 
into two categories, and the binary parameters $(s, q)$ in the first category were searched 
for using a grid approach with multiple starting values of $\alpha$, and the other parameters 
were found using a downhill approach based on the Markov Chain Monte Carlo (MCMC) method with 
an adaptive step size Gaussian sampler \citep{Doran2004}.  We then constructed a $\Delta\chi^2$ 
map on the $s$--$q$ parameter plane and identified local minima on the map. In the second stage, 
we refined the local solutions by allowing all parameters to vary, and then found a global 
solution by comparing the goodness of the fits of the individual local solutions.

From the 2L1S modeling, we found a pair of solutions with a projected binary-lens separation 
$s < 1$ (close solution) and a separation $s > 1$ (wide solution) resulting from the close--wide 
degeneracy \citep{Griest1998, Dominik1999, An2005}. The binary parameters are $(s, q)_{\rm close} 
\sim   (0.73, 2.2\times 10^{-3})$ for the close solution, and $(s, q)_{\rm wide}\sim  (1.37, 2.1
\times 10^{-3})$ for the wide solution. The full lensing parameters of the individual solutions 
are listed in Table~\ref{table:one} together with the $\chi^2$ values of the fits of the models.  
For both solutions, the estimated very low mass ratio $q\sim 2.1\times 10^{-3}$ between the lens 
components indicates that the companion to the lens is a planetary-mass object.  The lens-system 
configuration of the wide 2L1S solution is presented in the inset of the top panel in 
Figure~\ref{fig:two}. The configuration shows that the central anomaly was produced by the source 
passage very close to lower left cusp of the central caustic induced by a planet.  We note that 
the configuration of the close solution is very similar to that of the wide solution.  Although 
the source did not cross the caustic, it passes within 1.5 source radii from the caustic cusp. 
Hence, the normalized radius $\rho = (1.49 \pm 0.03)\times 10^{-3}$ was precisely measured from 
the peak part of the light curve deformed by finite-source effects.

\begin{figure}[t]
\includegraphics[width=\columnwidth]{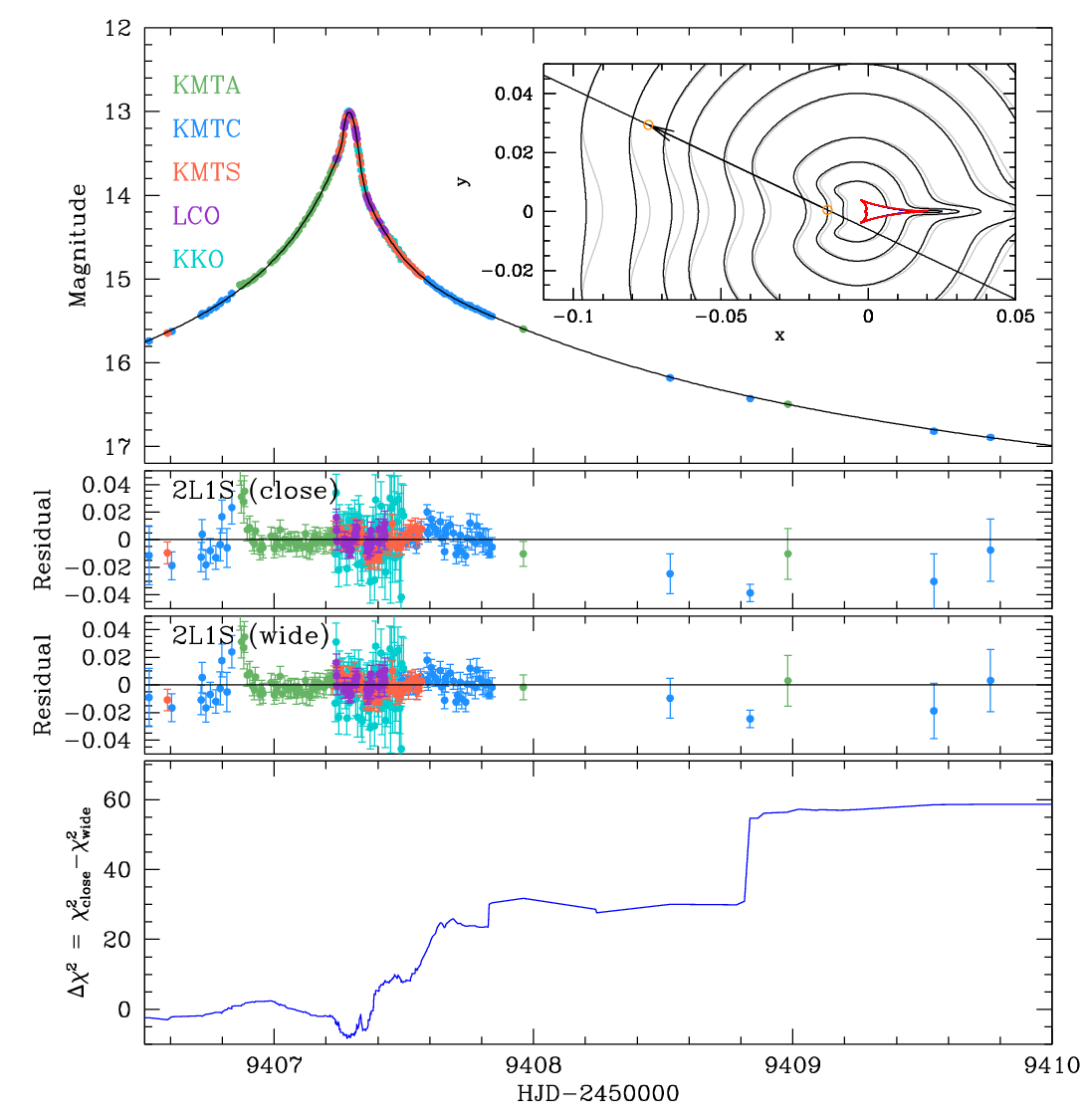}
\caption{
The cumulative diagram of $\Delta\chi^2=
\chi^2_{\rm close}-\chi^2_{\rm wide}$ between the close and wide 2L1S solutions (bottom panel).
The top panel shows the light curve in the same time range, and the two lower panels are the 
residuals from the two solutions.  The inset of the top panels shows the contour maps of lensing 
magnifications for the close (grey contours) and wide (black contours) solutions.
}
\label{fig:three}
\end{figure}

It was found that the wide solution is preferred over the close solution by $\Delta\chi^2=60.1$ 
despite that the two solutions are subject to the close-wide degeneracy.  In order to investigate 
the region of the fit difference, we present the cumulative diagram of $\Delta\chi^2=
\chi^2_{\rm close}-\chi^2_{\rm wide}$ between the two solutions in Figure~\ref{fig:three}.  The 
diagram shows that the wide solution yields better fits than than the close solution  in the two 
regions around ${\rm HJD}^\prime\sim 9407.5$ and $\sim 9408.8$.  In the inset of the top panel, we 
compare the contour maps of lensing magnifications for the close (grey contours) and wide (black 
contours) solutions.  It is found that the maps exhibit subtle differences despite the similarity 
between the caustics of the two solutions.  From this difference together with the the large number 
of data points contributing to $\chi^2$, the degeneracy between the close and wide solutions is 
lifted.

Figure~\ref{fig:two} shows the model curve (dotted curve in the top panel) and residual of the 
wide 2L1S solution.  The 2L1S model appears to approximately describe the anomaly, but detailed 
inspection reveals that the model leaves residuals from the model.  The most conspicuous residual 
appears at around the bump centered at $t_{\rm bump}\sim 9406.85$, while small but systematic 
negative residuals appear in the KMTA data in the part of the light curve after the bump during 
$9406.92 \lesssim  {\rm HJD}^\prime \lesssim  9407.2$. The bump in the residual is likely to be 
of astrophysical origin rather than systematics in the data, because it appears in 3 different 
data sets: KMTC, MOA, and KMTA. Despite the fact that 2L1S models that are very similar to ours 
were circulated around the microlensing community during the season of the event, an analysis of 
the event has not been published because the residual could not be fully explained with a 2L1S 
model. In order to explain the residual, we inspect more sophisticated models to check whether 
the residuals may vanish with other interpretations of the lens system.

\subsection{3L1S model }\label{sec:three-two}

In order to explain the residual from the 2L1S model, we check a model with a 3L1S
configuration by introducing an extra lens component to the binary-lens system. The consideration
of an extra lens component $M_3$ requires one to include additional lensing parameters in modeling.
These parameters are $(s_3, q_3, \psi)$, which denote the mass ratio and normalized projected 
separation between $M_1$ and $M_3$, and the orientation of $M_3$ as measured from the $M_1$--$M_2$ 
axis, respectively.  We use the notations $(s_2, q_2)$ for the parameters related to $M_1$--$M_2$ 
pair to distinguish them from those related to $M_3$. Because the 2L1S model approximately described 
the overall feature of the anomaly, we started searches for the third-body parameters $(s_3, q_3, 
\psi)$ via a grid approach by fixing the other lensing parameters as those of the 2L1S solution, 
and then refined the solution by allowing all parameters to vary. We carried out this procedure 
two times based on the close and wide 2L1S solutions.

\begin{table}[t]
\small
\caption{Model parameters of 3L1S solutions\label{table:two}}
\begin{tabular*}{\columnwidth}{@{\extracolsep{\fill}}lcccc}
\hline\hline
\multicolumn{1}{c}{Parameter}    &
\multicolumn{1}{c}{Close}        &
\multicolumn{1}{c}{Wide}        \\
\hline
$\chi^2$/dof            &   $1205.9             $   &  $1146.0             $     \\
$t_0$ (HJD$^\prime$)    &   $9407.280 \pm 0.001 $   &  $9407.278 \pm 0.001 $     \\
$u_0$ ($10^{-3}$)       &   $5.56 \pm 0.11      $   &  $5.48 \pm 0.10      $     \\
$\te$ (days)            &   $19.60 \pm 0.33     $   &  $20.24 \pm 0.34     $     \\
$s_2$                   &   $0.725 \pm 0.001    $   &  $1.372 \pm 0.002    $     \\
$q_2$ ($10^{-3}$)       &   $2.16 \pm 0.04      $   &  $2.12 \pm 0.04      $     \\
$\alpha$ (rad)          &   $0.451 \pm 0.002    $   &  $0.440 \pm 0.002    $     \\
$s_3$                   &   $1.012 \pm 0.014    $   &  $1.001 \pm 0.014    $     \\
$q_3$ ($10^{-6}$)       &   $1.8 \pm 0.2        $   &  $1.5 \pm 0.2        $     \\
$\psi$ (rad)            &   $5.528 \pm 0.005    $   &  $5.532 \pm 0.005    $     \\
$\rho$ ($10^{-3}$)      &   $1.54 \pm 0.03      $   &  $1.50 \pm 0.03      $     \\
\hline
\end{tabular*}
\end{table}

The lensing parameters of the 3L1S solutions found based on the close and wide 2L1S solutions 
are listed in Table~\ref{table:two}. For both solutions, the parameters related to $M_2$ are 
very similar to those of the 2L1S solutions, and the parameters related to $M_3$ are $(s_3, 
q_3)_{\rm close}\sim  (1.012, 1.8\times 10^{-6})$ and $(s_3, q_3)_{\rm wide}\sim  (1.001, 
1.5\times 10^{-6})$ for the close and wide solutions, respectively. These parameters indicate 
that the lens would be a two-planet system, in which the second planet has an extremely low 
planet-to-host mass ratio of order $10^{-6}$ and lies very close to the Einstein ring of the 
planet host.  If the signal of the second planet is real, then the measured mass ratio would 
be the lowest among the microlensing planets that have ever been detected.  Similarly to the 
case of the 2L1S solutions, it is found the wide solution yields a better fit than the close 
solution by $\Delta\chi^2=59.9$.

In the inset of the top panel in Figure~\ref{fig:two}, we present the lens-system configuration 
of the wide 3L1S solution. It shows that the second planet induces an additional caustic 
elongated along the $M_1$--$M_3$ axis, and the source passed through this caustic. This 
diminishes the residuals at around and after the bump at $t_{\rm bump}$, as shown in the 
residual of the wide 3L1S model presented in Figure~\ref{fig:two}. It is found that the 
introduction of the second planet improves the model fit by $\Delta\chi^2=42.4$ with respect 
to the 2L1S model.

\subsection{2L2S model }\label{sec:three-three}

It is known that a subtle deviation in a planetary signal can arise not only by an extra 
companion to the lens but also by a companion to the source as demonstrated in the case 
of the lensing event OGLE-2019-BLG-0304 \citep{Han2021b}.  Therefore, we additionally tested 
a 2L2S configuration of the lens system, in which an extra source was introduced to the 2L1S 
system. As in the case of the 3L1S modeling, the introduction of the source companion ($S_2$) 
to the primary source ($S_1$) requires one to include additional parameters. These parameters 
are $(t_{0,2}, u_{0,2}, \rho_2, q_F)$, which represent the time and separation at the moment 
of the closest $S_2$ approach to the lens, the normalized source radius of $S_2$, and the flux 
ratio between $S_1$ and $S_2$, respectively. We use the notations $(t_{0,1}, u_{0,1}, \rho_1)$ 
for the parameters related to $S_1$ to distinguish them from the parameters related to $S_2$.  
In the 2L2S modeling, the solution was found by testing various trajectories of the source 
companion based on the 2L1S solution considering the locations and amplitudes of the anomaly 
features that could not be fully explained by the 2L1S model.

\begin{figure}[t]
\includegraphics[width=\columnwidth]{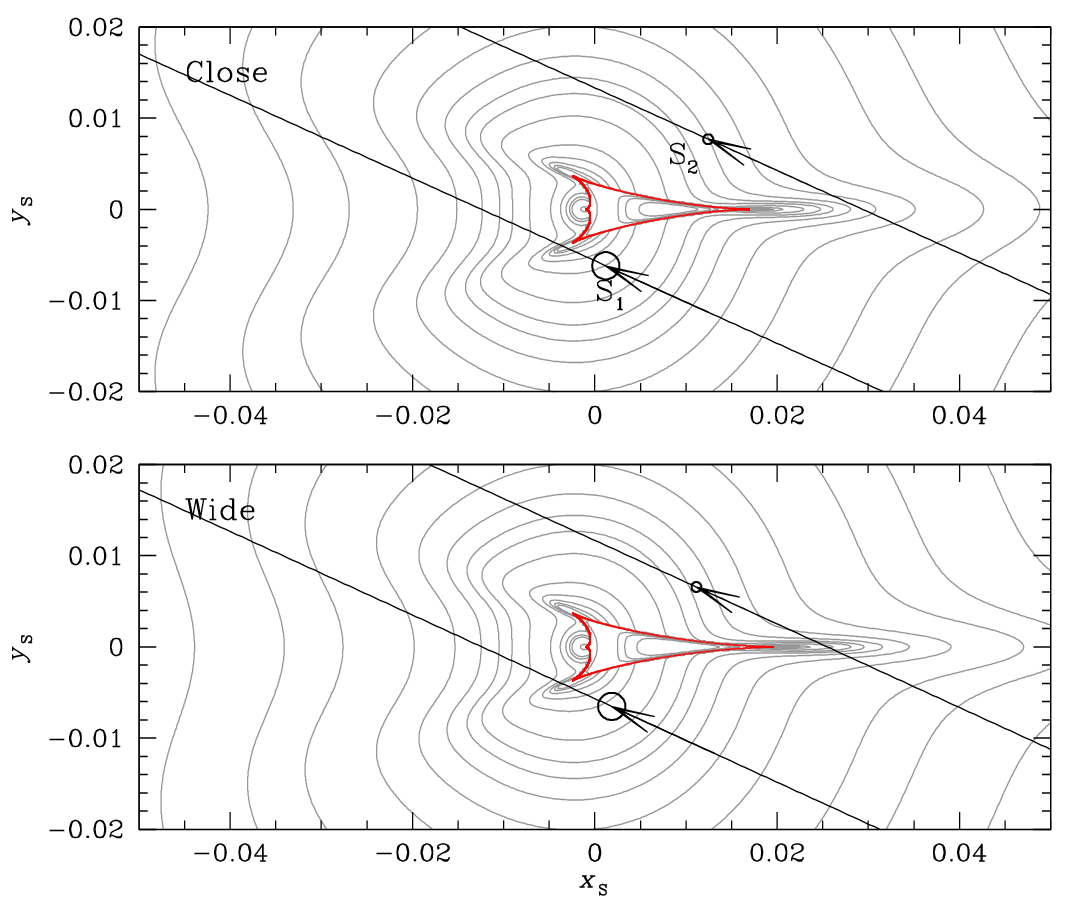}
\caption{
Lens-system configurations of the close and wide 2L2S solutions. In each panel, the red figure 
is the caustic, and the lines with arrows represent the trajectories of the primary (marked by 
"$S_1$") and secondary (marked by "$S_2$") source stars. The small empty circles on the source 
trajectories indicate the scaled sizes of the source stars. Grey curves encompassing the 
caustic represent equi-magnification contours.
}
\label{fig:four}
\end{figure}

In Table~\ref{table:three}, we list the lensing parameters of the close and wide 2L2S 
solutions found based on the close and wide 2L1S solutions, respectively.  Between the 
two solutions, it is found that the wide solution yields a better fit than the fit of the 
close solution by $\Delta\chi^2 = 19.0$. From the comparison of the parameters of the wide 
solution related to $S_1$, $(t_{0,1}, u_{0,1}) = (9407.277, 5.16\times 10^{-3})$, with 
those related to $S_2$, $(t_{0,2}, u_{0,2}) = (9407.339, -10.62\times 10^{-3})$, it is found 
that the secondary source passed on the opposite side of the primary source with respect to 
the planet host, trailing the primary with a slightly larger impact parameter than that of 
the primary source.  The flux ratio of between the source stars is $q_F \sim 5.7\%$ for the 
close solution and $\sim 2.6\%$ for the wide solution, indicating that the source companion 
is much fainter than the primary source. The lens-system configurations of the close and wide 
2L2S solutions are presented in the upper and lower panels of Figure~\ref{fig:four}, 
respectively.

\begin{table}[t]
\small
\caption{Model parameters of 2L2S solutions \label{table:three}}
\begin{tabular*}{\columnwidth}{@{\extracolsep{\fill}}lcccc}
\hline\hline
\multicolumn{1}{c}{Parameter}    &
\multicolumn{1}{c}{Close}        &
\multicolumn{1}{c}{Wide}        \\
\hline
 $\chi^2$                  &  $1140.3             $     &  $1121.3              $    \\
 $t_{0,1}$ (HJD$^\prime$)  &  $9407.278 \pm 0.001 $     &  $9407.277 \pm 0.001  $    \\
 $u_{0,1}$($10^{-3}$)      &  $5.15 \pm 0.10      $     &  $5.16 \pm 0.08       $    \\
 $\te$ (days)              &  $20.08 \pm 0.30     $     &  $20.90 \pm 0.29      $    \\
 $s$                       &  $0.734 \pm 0.001    $     &  $1.362 \pm 0.002     $    \\
 $q$ ($10^{-3}$)           &  $1.94 \pm 0.04      $     &  $1.95 \pm 0.03       $    \\
 $\alpha$ (rad)            &  $0.426 \pm 0.004    $     &  $0.430 \pm 0.003     $    \\
 $\rho_1$ ($10^{-3}$)      &  $1.49 \pm 0.02      $     &  $1.44 \pm 0.02       $    \\
 $t_{0,2}$ (HJD$^\prime$)  &  $9407.368 \pm 0.016 $     &  $9407.339 \pm 0.020  $    \\
 $u_{0,2}$ ($10^{-3}$)     &  $-12.13 \pm 0.43    $     &  $-10.62 \pm 0.51     $    \\
 $\rho_2$ ($10^{-3}$)      &  $0.54 \pm 0.28      $     &  $0.56 \pm 0.19       $    \\
 $q_F$                     &  $0.057 \pm 0.006    $     &  $0.026 \pm 0.004     $    \\
\hline
\end{tabular*}
\end{table}

\begin{figure}[t]
\includegraphics[width=\columnwidth]{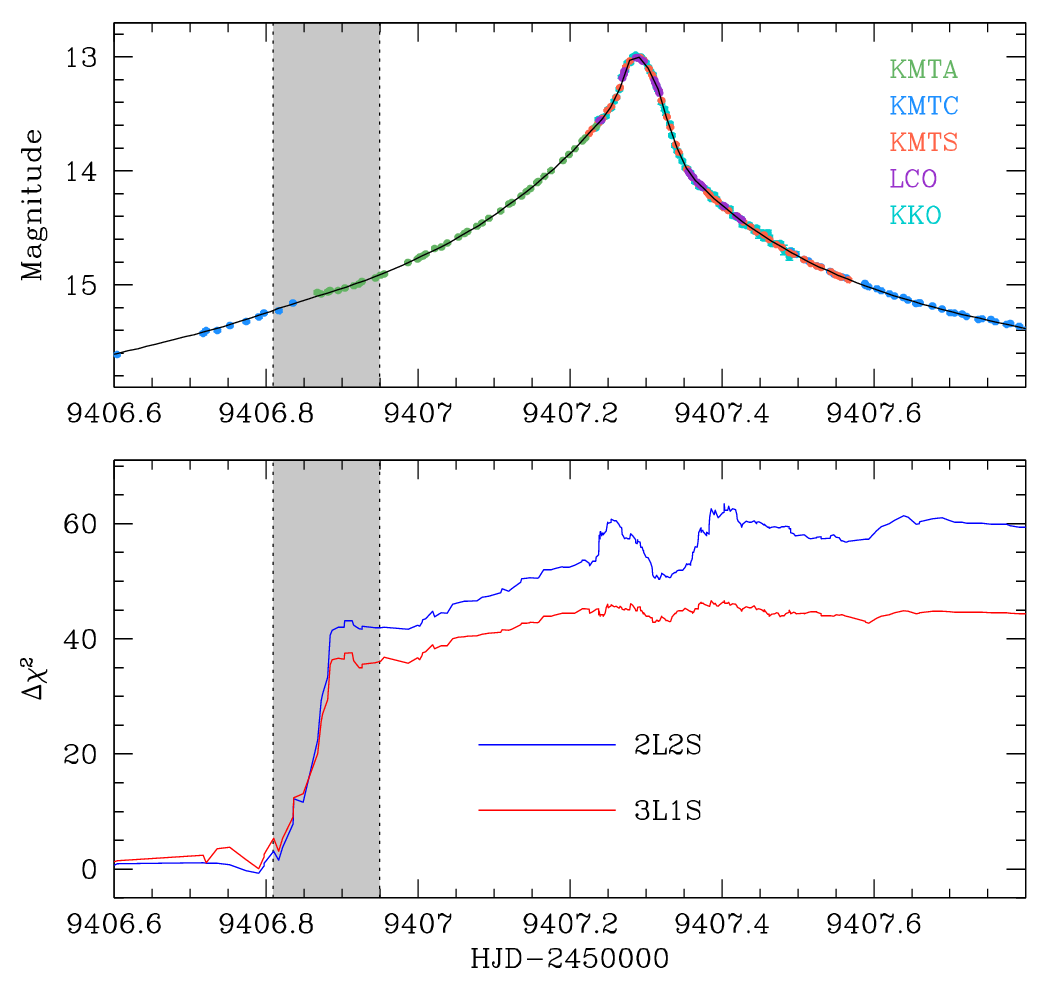}
\caption{
Cumulative $\chi^2$ distributions.  The blue curve is the distribution of $\chi^2$ difference 
between 2L1S and 2L2S solutions, $\Delta\chi^2_{\rm 2L2S}=\chi^2_{\rm 2L1S}- \chi^2_{\rm 2L2S}$, 
and the red curve represents the distribution of $\chi^2$ difference between the 3L1S and 2L1S 
solutions, $\Delta\chi^2_{\rm 3L1S}=\chi^2_{\rm 2L1S}- \chi^2_{\rm 3L1S}$.  The light curve in 
the upper panel is presented to show the region of fit improvement. The shaded region indicates 
the region of major fit improvement. 
}
\label{fig:five}
\end{figure}

We find that the 2L2S solution yields the best fit to the observed light curve among the three 
sets of tested models. From the comparison of the model fits, we find that the 2L2S solution yields 
a better fit than the 2L1S and 3L1S solutions by $\Delta\chi^2 = 67.1$ and 24.7, respectively. In 
Figure~\ref{fig:two}, we draw the model curve of the wide 2L2S solution over the data points and 
present the residuals from the model. From the inspection of the residuals, it is found that the 
residual bump at around $t_{\rm bump}$ vanishes and the negative 2L1S residuals during the period 
$9406.92 \lesssim {\rm HJD}^\prime \lesssim 9407.2$ substantially diminishes.  In Figure~\ref{fig:five}, 
we present the cumulative distributions of $\Delta\chi^2_{\rm 2L2S}=\chi^2_{\rm 2L1S}- \chi^2_{\rm 2L2S}$ 
(blue curve in the lower panel) and $\Delta\chi^2_{\rm 3L1S}=\chi^2_{\rm 2L1S}- \chi^2_{\rm 3L1S}$ (red 
curve) to show the region of fit improvement from the 2L1S model.  The distributions show that the fit 
improvement of the 2L2S model occurs throughout the anomaly region, while the improvement of the 3L1S 
model is mostly confined to the region around $t_{\rm bump}$.

\section{Source star and Einstein radius}\label{sec:four}

In this section, we specify the source stars of the event and estimate the angular Einstein 
radius of the lens system. The source stars were specified by measuring their dereddened colors 
and magnitudes, and the Einstein radius was estimated from the relation
\begin{equation}
\thetae = {\theta_* \over \rho},
\label{eq1}
\end{equation}
where the angular source radius $\theta_*$ was deduced from the source type, and the normalized 
source radius $\rho$ was measured from the modeling. In estimating $\thetae$, we use the angular 
and normalized source radii of the primary source, that is, $\thetae =  \theta_{*,1}$/$\rho_1$, 
because the uncertainties of $\theta_{*,1}$ and $\rho_1$ are much smaller than those of the 
secondary source star.

\begin{figure}[t]
\includegraphics[width=\columnwidth]{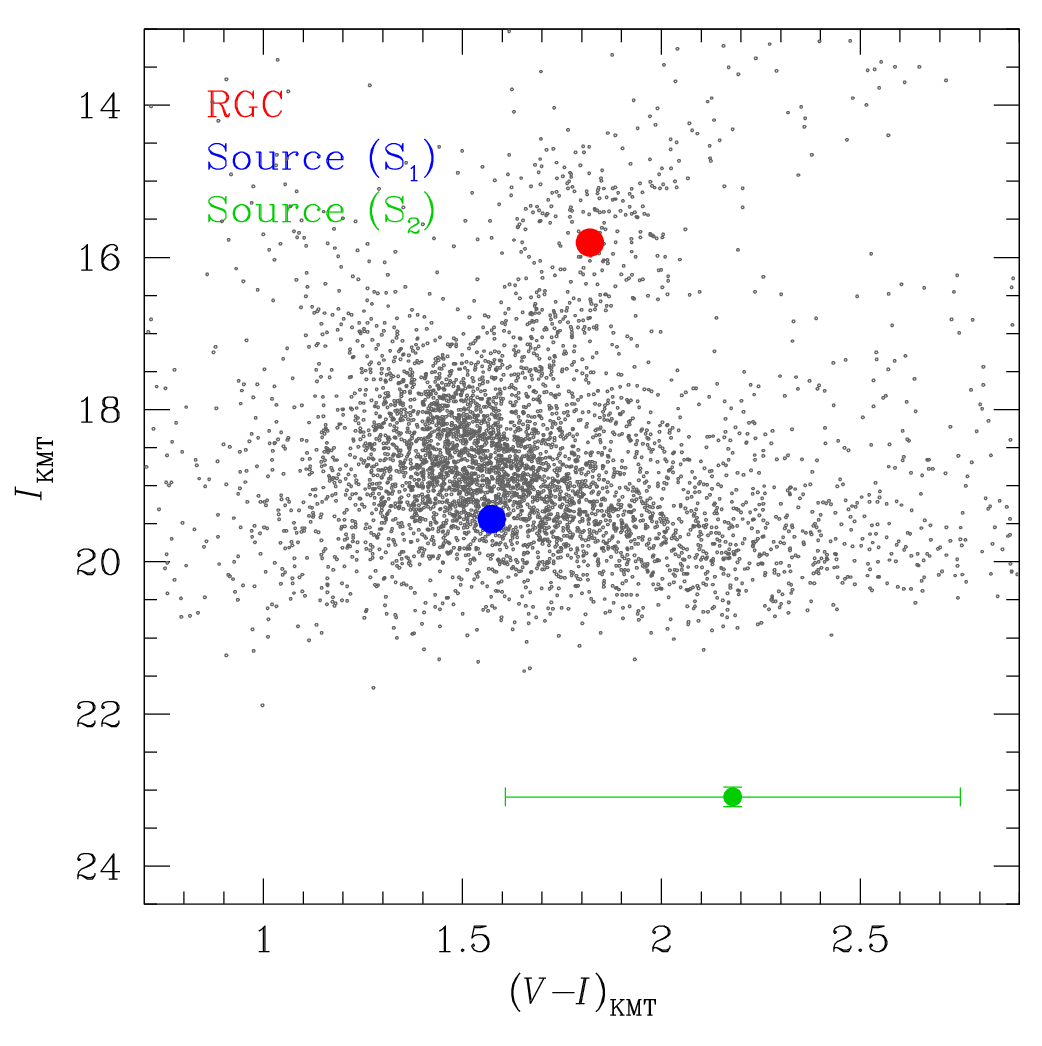}
\caption{
Locations of the primary and companion stars of the binary source with respect to the red 
giant clump (RGC) in the instrumental color-magnitude diagram of stars lying in the neighborhood 
of the source. 
}
\label{fig:six}
\end{figure}

For the measurements of the dereddened source color and magnitude, $(V-I, I)_0$, we first 
estimated the instrumental magnitudes of the source in the $I$ and $V$ bands, $(V, I)$, by 
regressing the photometric data processed using the pyDIA code with respect to the lensing 
model. We then assessed the flux values from the primary and secondary source stars, $(F_1, 
F_2)$, in each passband as
\begin{equation}
F_1 = {1\over 1+q_F}F_{\rm tot};\qquad
F_2 = {q_F\over 1+q_F}F_{\rm tot},
\label{eq2}
\end{equation}
where $F_{\rm tot} = F_1 + F_2$ is the combined source flux measured from the model regression, 
and the flux ratio between the binary source stars, $q_F = F_2/F_1$, is estimated from the 
modeling. Figure~\ref{fig:six} shows the locations of $S_1$ and $S_2$ in the instrumental 
color-magnitude diagram (CMD) of stars lying near the source constructed from the pyDIA 
photometry of these stars.  The measured instrumental color and magnitude are $(V-I, I)_1 = 
(1.574 \pm 0.012, 19.439 \pm 0.005)$ for the primary source and $(V-I, I)_2 = (2.179 \pm 
0.571, 23.088 \pm 0.130)$ for the secondary source.

We calibrated the instrumental source color and magnitude with the use of the \citet{Yoo2004}
routine, in which the centroid of the red giant clump (RGC), with $(V-I, I)_{\rm RGC} = (1.820, 
15.806)$ in the instrumental CMD, was used as a reference for calibration. With the measured 
offsets in color and magnitude of the source from the RGC centroid, $\Delta (V-I, I) = (V-I, I) 
- (V-I, I)_{\rm RGC}$, we estimated the dereddened values of the source as
\begin{equation}
(V-I, I)_0 = (V-I, I)_{{\rm RGC,0}} + \Delta(V-I, I), 
\label{eq3}
\end{equation}
where $(V-I, I)_{{\rm RGC},0} = (1.060, 14.322)$ represent the dereddened color and magnitude 
of the RGC centroid known from \citet{Bensby2013} and \citet{Nataf2013}, respectively. The 
estimated dereddened source color and magnitude from this procedure are
\begin{equation}
\eqalign{
(V-I, I)_{1,0} = (0.813 \pm 0.012, 17.955 \pm 0.005) & ~{\rm for}~S_1, \cr
(V-I, I)_{2,0} = (1.419 \pm 0.571, 21.604 \pm 0.130) & ~{\rm for}~S_2, \cr
}
\label{eq4}
\end{equation}
respectively.  According to the estimated colors and magnitudes, it is found that the primary 
source of the event is a subgiant of a late G or an early K spectral type, and the companion 
is a main-sequence star of a late K spectral type.

For the estimation of the source radius, we first converted $V-I$ 
color into $V-K$ color using the \citet{Bessell1988} relation, and then deduced the source 
radius from the relation between $\theta_*$ and $(V-K, I)$ of \citet{Kervella2004}. This 
yields the radii of the primary and secondary source stars of $\theta_{*,1} = 0.91 \pm 
0.06~\mu{\rm as}$ and $\theta_{*,2} = 0.30 \pm 0.17~\mu{\rm as}$, respectively. Finally, 
the Einstein radius was estimated using the relation in Eq.~(\ref{eq1}) as
\begin{equation}
\thetae = {\theta_{*,1}\over \rho_1} = 0.63 \pm 0.04~{\rm mas},
\label{eq5}
\end{equation}
and the relative lens-source proper motion was estimated as 
\begin{equation}
\mu =  {\thetae \over \te} = 11.02 \pm 0.79~{\rm mas}~{\rm yr}^{-1}.
\label{eq6}
\end{equation}
The values derived from $\theta_{*,2}$ are consistent, but significantly less precise than 
those above.  We inspected the Gaia data archive \citep{Gaia2018} to check the binarity of 
the source using the value of the Gaia Renormalized Unit Weight Error (RUWE).  The RUWE value 
is close to unity for a well-behaved single star solution, and a high value suggests the 
binarity of stars.  From this inspection, we found that the source is not registered in the 
Gaia archive, and thus it was difficult to check the source binarity based on the Gaia data.

We note that the estimated value of the relative lens-source proper motion may be subject 
to an additional uncertainty caused by the internal motion of the source induced by the 
source orbital motion.  According to the wide 2L2S solution, the projected separation between 
the component stars of the binary source, $\Delta\theta_{s,\perp}$, in units of the primary 
source is 
\begin{equation}
{\Delta\theta_{s,\perp}  \over \theta_{*,1} } =
{\Delta u \over \rho_1} =
{ [\Delta u_0^2 + (\Delta t_0/\te)^2]^{1/2}\over \rho_1} 
\sim 11.1,
\label{eq7}
\end{equation}
where $\Delta u_0=|u_{0,2}-u_{0,1}|$ and $\Delta t_0=|t_{0,2}-t_{0,1}|$ denote the differences 
between the impact parameters and closest approach times of the $S_1$ and $S_2$ trajectories, 
respectively. With $\theta_{*,1}\sim 0.91~\mu{\rm as}$ together with the adopted distance to 
the source of $\ds = 8$~kpc, the projected physical separation between the source stars is 
\begin{equation}
a_{\perp, s} = \ds \left( {\Delta u \over \rho_1} \right)\theta_{*,1}\sim 0.081~{\rm AU}.
\label{eq8}
\end{equation}
By adopting the primary source mass of $M_{s,1} = 1~M_\odot$ and the secondary source mass 
of $M_{s,2}=0.6~M_\odot$, and assuming a circular face-on orbit, this yields a source orbital 
period of $P \sim 6.65$~days. Then the internal velocity of the binary-source system would be 
$v_{\rm int} = 30~{\rm m}~{\rm s}^{-1}\times  (a_{\perp,s}/{\rm AU})/(P/{\rm yr}) = 
134~{\rm km}~{\rm s}^{-1}$, which corresponds to the internal proper motion
\begin{equation}
\mu_{\rm int} = {v_{\rm int} \over \ds} = 2.5~{\rm mas}~{\rm yr}^{-1},
\label{eq9}
\end{equation}
This internal proper motion is a non-negligible fraction of the proper motion $\mu =
11.02~{\rm mas}~{\rm yr}^{-1}$ estimated without considering the internal source motion.

The internal motion of the source can have several effects. First, the normalized source 
radius of the source companion, $\rho_2$, can be somewhat different from the value that is 
found from the model fit. The internal proper motion of $S_2$ relative to the center of mass 
would be $\mu_{\rm int,2} = [M_{S,1}/(M_{S,1}+M_{S,2})] \mu_{\rm int}\sim 
2.2~{\rm mas}~{\rm yr}^{-1}$. Then, $\rho_2$ could be a factor $(1 \pm \mu_{\rm int,2}/\mu) 
= 1 \pm 0.2$ larger or smaller than what is found in the fit. However, we note that this 
does not qualitatively affect the result because the uncertainty of $\rho_2$ is already very 
big. Second, the primary would move relative to the center of mass with a proper motion 
$\mu_{\rm int,1} = [M_2/(M_1+M_2)] \mu_{\rm int}\sim 1.3~{\rm mas}~{\rm yr}^{-1}$, which is 
about 12\% of the value $\mu =11.02~{\rm mas}~{\rm yr}^{-1}$ obtained without considering the 
internal motion. This implies that the estimated proper motion is subject to an additional 
$\sim 12\%$ uncertainty due to internal proper motion, which should be considered when future 
adaptive optics observations are made.

\begin{figure}[t]
\includegraphics[width=\columnwidth]{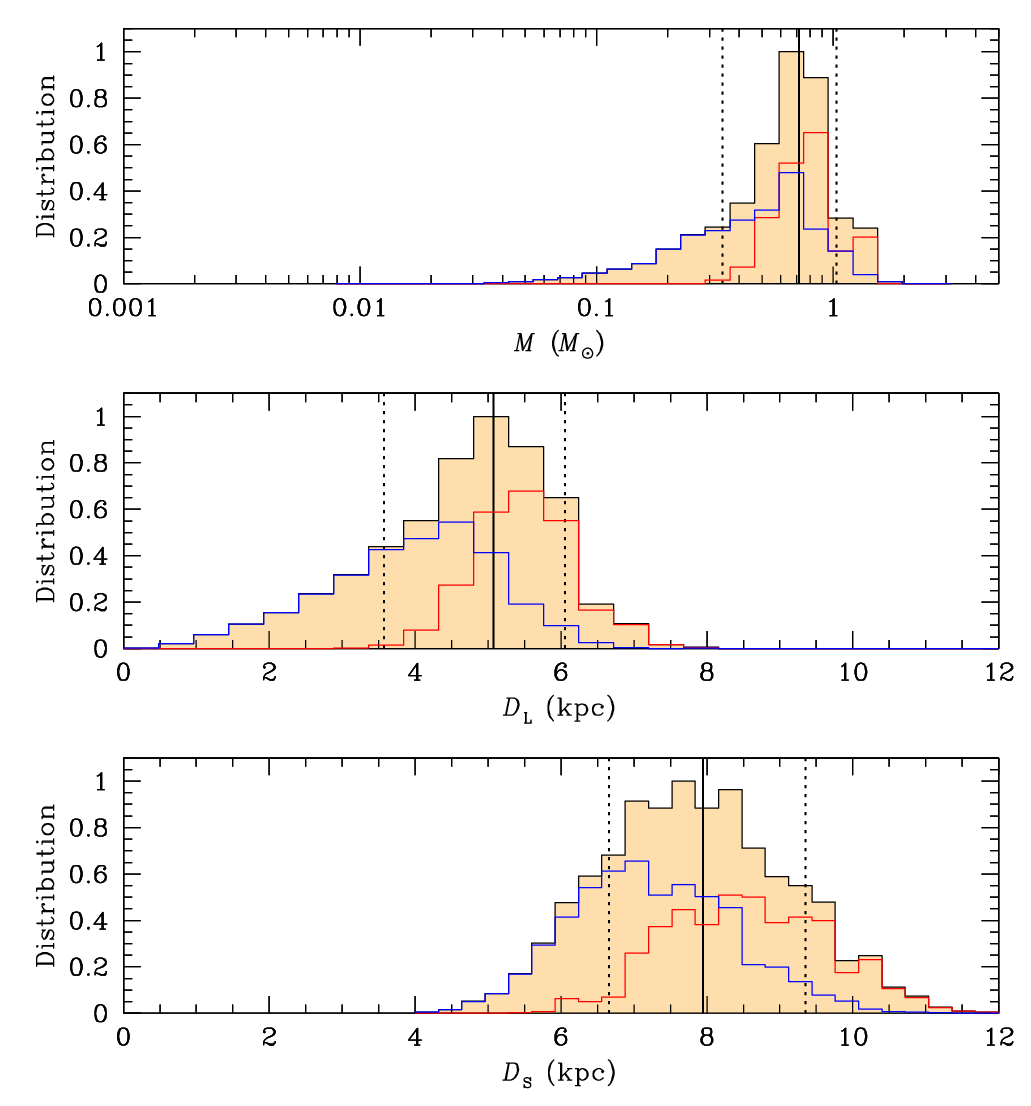}
\caption{
Bayesian posteriors of the primary lens mass and distance to the lens and source.  In each 
panel, the solid vertical line represents the median value, and the two dotted vertical 
lines indicate the 16\% and 84\% of the posterior distribution. The blue and red curves 
present the contributions by the disk and bulge lens populations, respectively, and black 
curve is the sum of the two lens populations. 
}
\label{fig:seven}
\end{figure}

\section{Physical parameters}\label{sec:five}

We estimate the physical parameters of the mass $M$ and distance $\dl$ to the planetary lens
system using the measured lensing observables of $\te$ and $\thetae$, which are related to the 
mass and distance to the lens by the relations
\begin{equation}
\te = {\thetae \over \mu};\qquad
\thetae = (\kappa M \pirel)^{1/2},
\label{eq10}
\end{equation}
respectively. Here $\kappa =4G/(c^2{\rm AU})\simeq 8.14~{\rm mas}~M_\odot^{-1}$ and $\pirel= 
{\rm AU}(D_{\rm L}^{-1} - D_{\rm S}^{-1})$  represents the relative lens-source parallax. The 
physical parameters can be uniquely determined with the additional observable of the microlens 
parallax $\pie$ by the relations 
\begin{equation}
M = {\thetae \over \kappa\pie};\qquad
\dl = { {\rm AU}\over \pie\thetae +\pi_{\rm S}}
\label{eq11}
\end{equation}
\citep{Gould2000},
but $\pie$ could not be securely measured for KMT-2021-BLG-1547 because of the relatively short
time scale, $\te \sim 21$~day, of the event. Therefore, we estimate $M$ and and $\dl$ by conducting 
a Bayesian analysis based on the measured observables of $\te$ and $\thetae$.

The Bayesian analysis was conducted with the use Galaxy and mass-function models. Based on 
these models, we produced a large number of artificial lensing events, for which the locations 
of the lens and source and their relative proper motions were derived from the Galactic model, 
and the lens mass were derived from the mass-function model from a Monte Carlo simulation. In 
this simulation, we used the \citet{Jung2021} Galaxy model and the \citet{Jung2018} mass-function 
model.  In the Galaxy model, the density profile of disk objects follows the modified 
double-exponential form presented in second line of Table~3 in \citet{Robin2003}, and the bulge 
profile follows the triaxial model of \citet{Han1995}.  The motion of disk objects follows the 
model constructed originally based on the \citet{Han1995} and modified to reconcile  the 
\citet{Robin2003} disk density profile.  The motion of bulge objects is modeled based on the 
proper motions of stars in the Gaia catalog \citep{Gaia2016, Gaia2018}.  The mass functions 
(MFs) of disk and bulge lens objects are constructed by adopting the initial and present-day 
MFs of \citet{Chabrier2003}, respectively.  For the individual artificial events, we computed 
the Einstein time scales and Einstein radii using the relations in Eq.~(\ref{eq10}), and then 
constructed the posteriors of the lens mass and distance by imposing a weight $w_i = 
\exp(-\chi^2/2)$ to each event.  Here we compute $\chi^2$ value as
\begin{equation}
\chi^2 = 
\left( {t_{{\rm E},i} -\te \over \sigma_{\te}}\right)^2 + 
\left( {\theta_{{\rm E},i} -\thetae \over \sigma_{\thetae}}\right)^2,
\label{eq12}
\end{equation}
where $(t_{{\rm E},i}, \theta_{{\rm E},i})$ are the time scale and Einstein radius of each 
simulated event, and $(\te, \thetae)$ and $(\sigma_{\te}, \sigma_{\thetae})$ represent the 
measured values and their uncertainties, respectively.

\begin{table}[t]
\small
\caption{Model parameters of 2L2S solutions \label{table:four}}
\begin{tabular*}{\columnwidth}{@{\extracolsep{\fill}}lcccc}
\hline\hline
\multicolumn{1}{c}{Parameter}        &
\multicolumn{1}{c}{Value}        \\ 
\hline 
 $M_{\rm h}$ ($M_\odot$)    &  $0.72^{+0.32}_{-0.38} $      \\  [0.8ex]
 $M_{\rm p}$ ($M_{\rm J}$)  &  $1.47^{+0.65}_{-0.77} $      \\  [0.8ex]
 $\dl$ (kpc)                &  $5.07^{+0.98}_{-1.50} $      \\  [0.8ex]
 $a_\perp$ (AU)             &  $4.5^{+0.9}_{-1.3}    $      \\  [0.5ex]
\hline
\end{tabular*}
\end{table}

In Figure~\ref{fig:seven}, we present the Bayesian posteriors of the primary lens mass, distance 
to the lens and source. In Table~\ref{table:four}, we list the estimated values of the host mass 
$M_{\rm h}$, planet mass $M_{\rm p}$, distance to the planetary system, and the projected separation 
between the planet and host, $a_\perp=s\thetae\dl$.  We use the median values of the posterior 
distributions as representative values, and the uncertainties were estimated as the 16\% and 84\% 
of the distributions. According to the estimated parameters, the lens is a planetary system, in 
which a planet with a mass about 50\% more massive than the Jupiter of the solar system orbits a 
host star with a mass about 30\% less massive than the sun. The projected planet--host separation 
$a_\perp \sim 4.5$~AU is substantially greater than the snow line $a_{\rm snow}\sim 2.7(M/M_\odot) 
\sim 1.9$~AU of the planetary system, indicating that the planet lies well beyond the snow line 
of the planetary system. In each posterior distribution, we mark the contributions by the disk 
(blue curve) and bulge (red curve) lens populations. It is found that the relative probabilities 
for the planet host to be in the disk and bulge are 55\% and 45\%, respectively

\section{Summary and discussion}\label{sec:six}

In our recent project, we have inspected the previous microlensing data collected by the 
KMTNet survey in search of anomalous lensing events for which there have been no suggested 
models precisely describing the observed anomalies. Following the analyses on the events 
OGLE-2018-BLG-0584 and KMT-2018-BLG-2119 by \citet{Han2023a} and on the KMT-2021-BLG-1122 
by \citet{Han2023b}, we conducted the analysis on the event KMT-2021-BLG-1547, for which 
the anomaly in the lensing light curve could not be precisely described by a usual 
binary-lens model.

We investigated the origin of the residuals by testing more sophisticated models that 
included either an extra lens component or an extra source star to the 2L1S configuration 
of the lens system.  From these analyses, we found that the residuals from the binary-lens 
model originated from the existence of a faint companion to the source. It was found that 
the 2L2S solution substantially diminished the residuals and improved the model fit by  
$\Delta\chi^2 = 67.1$ with respect to the 2L1S solution. It was found that the 3L1S 
solution also improved the fit, but the fit was worse than that of the 2L2S solution by 
$\Delta\chi^2 = 24.7$.

An important scientific goal of the microlensing surveys is to reveal the demographic 
properties of extrasolar planets, especially those lying in the outer regions of planetary 
systems. For such studies, it is important to accurately assess the detection efficiency 
that is based on a complete planet sample constructed under well-defined detection criteria. 
If a fraction of planets were missed in this sample because their signals cannot be fully 
explained, this would lead to erroneous estimation of the detection efficiency, and thus 
incorrect results on the demographic properties. The event KMT-2021-BLG-1547 demonstrates 
the need of sophisticated modeling for unexplained anomalies for the construction of a 
complete microlensing planet sample.

\begin{acknowledgements}
Work by C.H. was supported by the grants of National Research Foundation of Korea 
(2020R1A4A2002885 and 2019R1A2C2085965).
This research has made use of the KMTNet system operated by the Korea Astronomy and Space Science 
Institute (KASI) at three host sites of CTIO in Chile, SAAO in South Africa, and SSO in Australia. 
Data transfer from the host site to KASI was supported by the Korea Research Environment Open NETwork 
(KREONET).
This research was supported by the Korea Astronomy and Space Science Institute under the R\&D
program (Project No. 2023-1-832-03) supervised by the Ministry of Science and ICT.
The MOA project is supported by JSPS KAKENHI
Grant Number JSPS24253004, JSPS26247023, JSPS23340064, JSPS15H00781,
JP16H06287, and JP17H02871.
J.C.Y., I.G.S., and S.J.C. acknowledge support from NSF Grant No. AST-2108414. 
Y.S.  acknowledges support from NSF Grant No. 2020740.
W.Zang acknowledges the support
from the Harvard-Smithsonian Center for Astrophysics through the CfA Fellowship. 
This research uses data obtained through the Telescope Access Program (TAP), which has been
funded by the TAP member institutes. W.Zang, H.Y., S.M., and W.Zhu acknowledge support by the
National Natural Science Foundation of China (Grant No. 12133005).
C.R. was supported by the Research fellowship of the Alexander von Humboldt Foundation.
\end{acknowledgements}

\end{document}